\documentclass[journal=nalefd,manuscript=letter]{achemso}
\setkeys{acs}{articletitle = true}

\usepackage{amsmath}
\usepackage{graphicx}
\usepackage{subcaption}
\usepackage[colorinlistoftodos]{todonotes}
\usepackage[colorlinks=true, allcolors=blue]{hyperref}
\usepackage{siunitx}
\usepackage{layouts}
\usepackage{soul} 
\usepackage{nameref} 

\title{Elastomeric waveguide on-chip coupling of an encapsulated MoS$_2$ monolayer}

\author{Filip Auksztol}
\email{oxon@nus.edu.sg}
\affiliation{Centre for Quantum Technologies, National University of Singapore}

\author{Daniele Vella}
\affiliation{Department of Physics, National University of Singapore}

\author{Ivan Verzhbitskiy}
\affiliation{Department of Physics, National University of Singapore}

\author{Kian Fong Ng}
\affiliation{Centre for Quantum Technologies, National University of Singapore}

\author{Yi Wei Ho}
\affiliation{NUS Graduate School for Integrative Sciences and Engineering}
\alsoaffiliation{Centre for Advanced 2D Materials and Graphene Research Centre}
\alsoaffiliation{Department of Physics, National University of Singapore}

\author{James A. Grieve}
\affiliation{Centre for Quantum Technologies, National University of Singapore}

\author{Jos\'e Viana-Gomes}
\affiliation{Department of Physics, National University of Singapore}
\alsoaffiliation{Centre for Advanced 2D Materials and Graphene Research Centre}

\author{Goki Eda}
\affiliation{Department of Physics, National University of Singapore}
\alsoaffiliation{Department of Chemistry, National University of Singapore}
\alsoaffiliation{Centre for Advanced 2D Materials and Graphene Research Centre}

\author{Alexander Ling}
\email{cqtalej@nus.edu.sg}
\affiliation{Centre for Quantum Technologies, National University of Singapore}
\alsoaffiliation{Department of Physics, National University of Singapore}

\begin{document}

\sisetup{range-phrase=--, range-units=single}

\maketitle

\begin{abstract}
We propose a robust photonic platform for encapsulation and addressing of optically active 2D- and nano-materials. Our implementation utilises a monolayer of MoS$_2$ transition metal dichalcogenide embedded in an elastomeric waveguide chip. The structure is manufactured from PDMS using soft-lithography and capable of sustaining a single mode of guided light. We prove that this setup facilitates addressing of the 2D material flake by pumping it with polarised laser light and gathering polarisation-resolved photoluminescence spectra with the extinction ratio of 31, which highlights the potential for selection-rule dependent measurements. Our results demonstrate improved handling of the material and experimental simplification compared to other techniques. Furthermore, inherent elasticity of the host provides an avenue for direct mechanical coupling to embedded materials.
\end{abstract}

\begin{figure}
\centering
\includegraphics{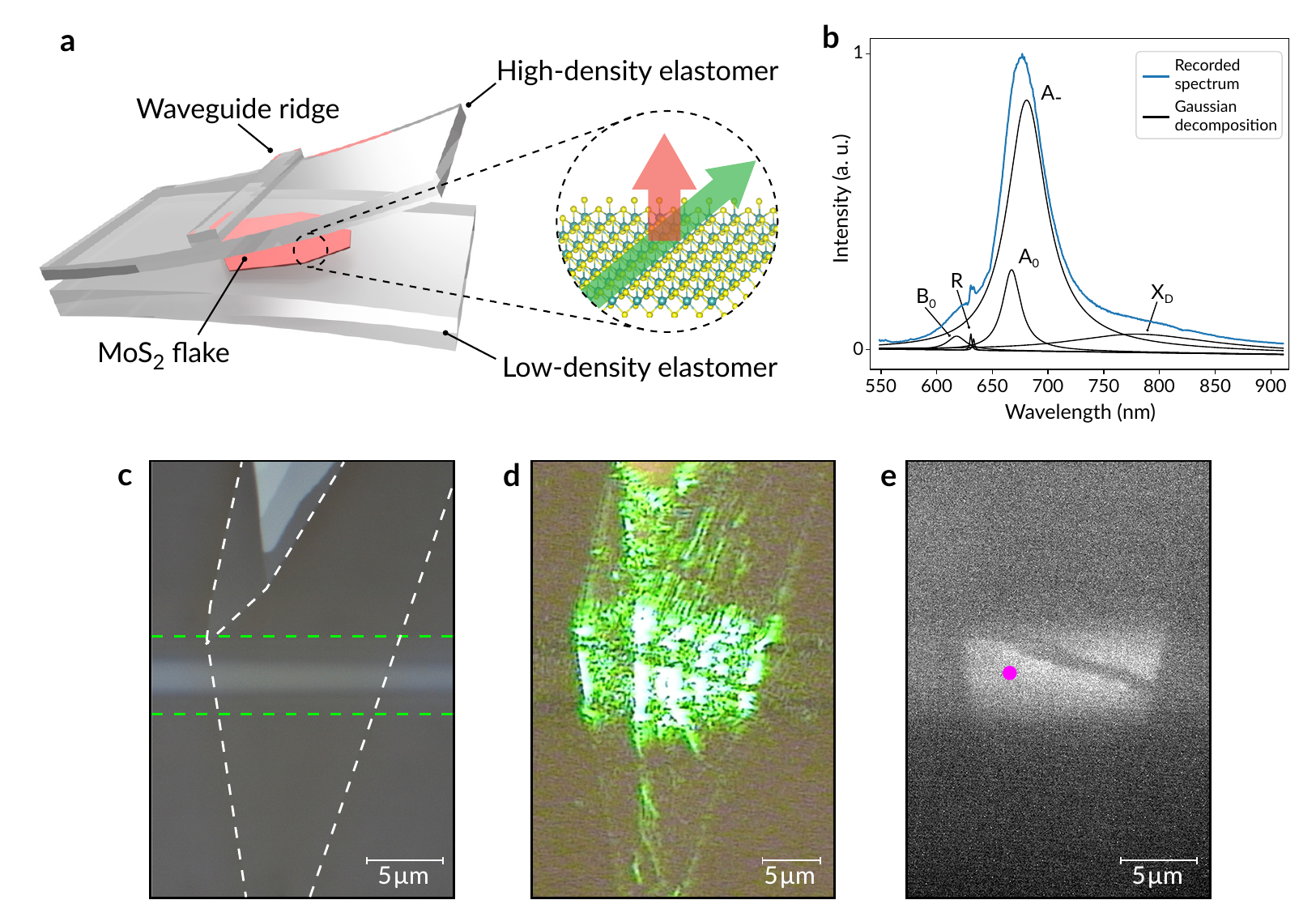}
\caption{\label{fig:panel} Elastomeric waveguide structure with optically active material encapsulated within the guided mode of light. (a) Artistic representation of the monolayer of MoS$_2$ encapsulated within the waveguide structure comprised of two layers of elastomer with different densities. Call-out shows the schematic representation of addressing (green arrow) and collection (red arrow) geometry with respect to the MoS$_2$ monolayer. (b) Waveguide excitation photoluminescence spectrum collected from the encapsulated MoS$_2$ monolayer. Physical origins of the individual components of the spectrum are deduced by fitting the Gaussian functions, which correspond to: neutral B-exciton (B\textsubscript{0}), substrate Raman doublet (R), neutral A-exciton (A\textsubscript{0}), charged A-exciton (A$_{-}$), defect emission (X\textsubscript{D}). (c) Optical microscope image ($\times100$ magnification) of the encapsulated flake. White dashed lines outline the flake and the boundary between the monolayer and bulk material. Green dashed lines outline the edges of the waveguide (\SI{5}{\micro\metre} width). (d) Flake illuminated with \SI{532}{\nano\metre} pump delivered to the sample via the waveguide. (e) Optical microscope image of the PL emission during the waveguide excitation of the sample. Pump is blocked with a long-pass filter. Magenta dot indicates the probing region, from which the PL spectrum has been collected during the polarisation dependence experiment.}
\end{figure}



Advances in the fundamental understanding of ultra-thin transition metal dichalcogenides (TMDs) including their excitonic properties \cite{robert2017}, exciton-phonon coupling \cite{niehues2018}, optical selection rules \cite{wang2017} and valley degree of freedom \cite{xu2014}, position them as attractive candidates for components of devices in novel photonic and optoelectronic applications \cite{vella2017,huo2017,hill2017,wu2014}. Monolayer materials are fragile and difficult to handle due to the micrometer footprint and their tendency to crumple. In order to incorporate them in the existing photonic systems, there is a need for packaging in the form of a robust platform for handling and testing of the material. Planar photonic waveguide chips capable of confining light on the micrometer scale are a natural candidate for such a platform, offering transmission of visible light, polarisation selectivity and shielding from environmental influence. While this is challenging to achieve with traditional semiconductor or glass waveguide structures, we propose that a photonic device fabricated in an elastomer can serve as a host for encapsulation of the monolayers and their heterostructures. Using the suggested geometry, we show it is feasible to confine an optically active 2D material within the mode of guided light, allowing for the propagation of linearly polarised light in the plane of the monolayer. As reported previously~\cite{grieve2017} such waveguides are capable of guiding orthogonal polarization states with very similar mode profiles. This allows the polarisation of light in this system to be changed such that exciton states with different optical selection rules (e.g. dark and bright excitons) can be independently accessed. Embedding the material of interest in an elastomer gives an added benefit of direct mechanical coupling by performing either global or local deformation of the chip.

Here, we describe a proof-of-concept device, utilising a monolayer of MoS$_2$ encapsulated within the mode of light guided in a polydimethylsiloxane (PDMS) chip. A stylised illustration of the experimental sample (Figure \ref{fig:panel}a) shows an MoS$_2$ flake inserted between the two layers of PDMS. Different refractive indices of the layers (1.4320 and 1.4311, simulated), together with the geometry of the top ridge, create a light-guiding structure capable of sustaining a single mode of radiation (see inset in Figure \ref{fig:setupPL} for an image of the guided mode). Positioning a flake of 2D material within the mode of the guided light allows for efficient coupling. Here, we demonstrate this experimentally for a flake of MoS$_2$ by exciting it through the waveguide and performing room temperature measurement of the photoluminescence spectrum while varying the polarisation of incident light.

\subsection{Experiment}

\begin{figure}
\centering
\includegraphics[width=\textwidth]{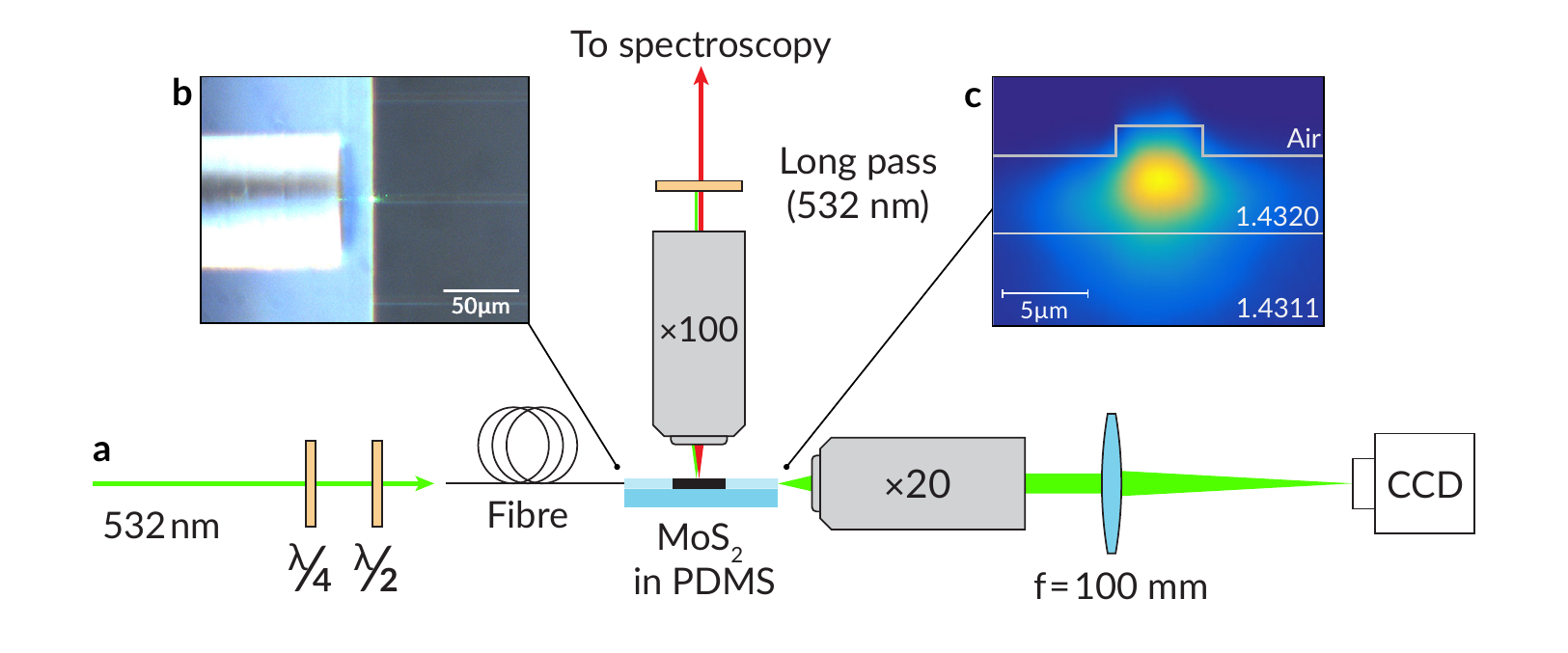}
\caption{\label{fig:setupPL} (a) Photoluminescence experimental setup (chip side-view perspective). Exposed fibre tip, the waveguide chip and the imaging microscope are each mounted independently on micrometre stages for alignment. The top-down microscope is part of a larger spectroscopy setup (custom-designed NTEGRA system by NT-MDT). (b) Microscope image of the fibre tip coupling \SI{532}{\nano\metre} pump laser to the waveguide chip. \SI{125}{\micro\metre} diameter cladding of the fibre is visible, together with three \SI{5}{\micro\metre} waveguides on the chip. (c) Experimental TEM mode of the transmitted light, measured at the end-face of the waveguide chip (see arrow).}
\end{figure}

The experimental setup for the excitation and collection of PL spectra is shown in Figure \ref{fig:setupPL}. \SI{532}{\nano\metre} laser light is coupled to a single-mode fibre, which  is cut and stripped at one end to expose the core and cladding. The exposed end is mounted on a micrometre stage (Newport ULTRAlign XYZ + yaw, pitch, roll) and aligned to couple to the chip waveguide that encapsulates the MoS$_2$ flake. Coupling is optimised by maximising the transmitted power collected by the imaging objective on the opposite end of the chip ($\times20$). The incident light couples to the MoS$_2$ monolayer and the resulting photoluminescence is collected by a high NA top-down objective ($\times100$), after passing through the long-pass filter to eliminate the excitation wavelength. To achieve linear polarization at the location of the MoS$_2$ flake, strain-induced birefringence in the single mode fibre (and to a lesser extent the waveguide itself) is pre-compensated using a quarter wave plate and half wave plate. Once completed, the polarization of the excitation beam at the location of the flake can be rotated in the linear basis by additional rotation of the half wave plate alone.

\subsection{Results}

\begin{figure}
\centering
\includegraphics[width=\textwidth]{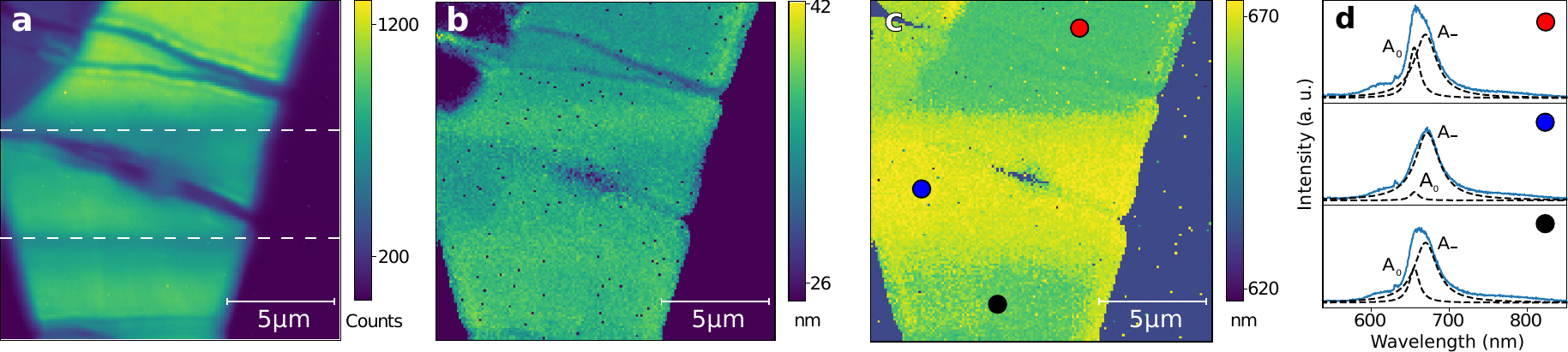}
\caption{\label{fig:PLMAp} Photoluminescence maps of the flake area obtained with NTEGRA (NT-MDT) piezo-controlled spectrometer. (a) Integrated photoluminescence in the vicinity of the waveguide ridge. White dashed lines indicate edges of the waveguide. Irregularities in the monolayer are visible in the waveguide path. (b) Linewidth broadening. (c) Peak position showing a strong correlation with waveguide geometry. (d) Sample spectra obtained from the approximate positions marked in (c). See text for explanation.}
\end{figure}

The encapsulated flake is investigated in detail by spatial mapping of the photoluminescence spectrum by normal excitation, which shows the distribution of the integrated photoluminescence, linewidth broadening and the peak position for the \SIrange{620}{720}{\nano\metre} spectral range (Figures \ref{fig:PLMAp}a, b and c respectively). Figure \ref{fig:PLMAp}a reveals the integrated photoluminescence spectrum map for the area of the flake within and around the edges of the waveguide ridge, with variations that can be attributed to the spatial inhomogeneity of the flake. This is likely due to wrinkles formed during the fabrication of the low-density elastomer layer of the waveguide. There is a correlation between the linewidth distribution and the photoluminescence intensity (compare Figure \ref{fig:PLMAp}a and b), with narrow emission implying high area and intensity.

The map of the peak distribution (Figure~\ref{fig:PLMAp}c) shows a strong correlation with the waveguide geometry, with the two regions of interest separated by the edges of the waveguide ridge. The photoluminescence peak undergoes a red shift of $\sim$\,\SI{17}{\nano\metre} between the outer region towards the waveguide ridge. Figure~\ref{fig:PLMAp}d shows extracted photoluminescence spectra from different regions of the monolayer flake. Such spectrum mainly consists of the ground exciton peak \cite{splendiani2010} originating from the direct optical transition at the \textrm{K} and \textrm{K'} points of the Brillouin zone. \cite{mak2010,chhowalla2013} The strong asymmetric photoluminescence peak is made of two contributions: neutral (A$_0$) and charged (A$_-$) excitons \cite{mak2013,chernikov2015} separated by ~ 45 meV (See Fig. 1b). Noticeably, relative intensities of these two peaks vary spatially, resulting in the apparent changes of the cumulative line-shape. Similar modifications of the optical properties have been reported for uniaxially \cite{niehues2018} and biaxially \cite{lloyd2016} strained monolayer of MoS$_2$.

\begin{figure}
\centering
\includegraphics[width=\textwidth]{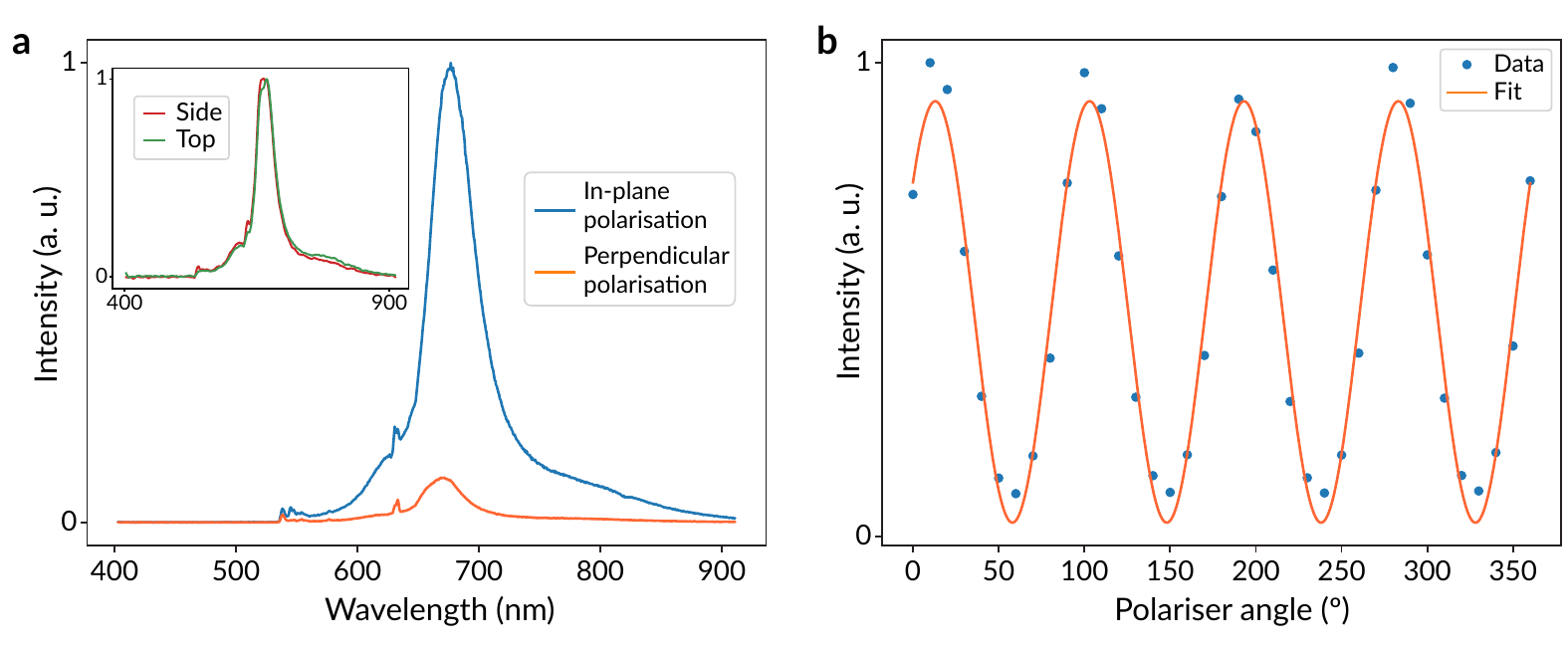}
\caption{\label{fig:PL} Photoluminescence measurement with \SI{532}{\nano\metre} pump. (a) Photoluminescence spectra of encapsulated MoS$_2$ flake excited through the wavegudie with incident in-plane polarisation (blue) and perpendicular polarisation (orange). Inset shows normalized spectra with top and side (through the waveguide with in-plane polarisation) excitation. (b) Photoluminescence spectra for intermediate polarisation states, integrated and normalised for power (blue) with sine function fit (orange). Extinction ratio of fitted function is 31 with oscillation period of \SI{90.0}{\degree} (which corresponds to an \SI{180.0}{\degree} rotation period of the polarisation).}
\end{figure}

Subsequently, we study the polarisation dependence of the photoluminescence spectra, which are obtained by parallel excitation of the flake with incident pump light delivered to the material through the waveguide (see call-out in Figure \ref{fig:panel}a). A monolayer of MoS$_2$ is pumped with in-plane \SI{532}{\nano\metre} laser light to induce the primary photoexcitations (excitons)  (Figure \ref{fig:PL}a). The resultant photoluminescence spectrum is collected from the probing region (Figure \ref{fig:panel}e) and reveals the typical characteristics of monolayer MoS$_2$.

Polarisation parallel to the flake surface produced maximum excitation (photoluminescence) and conversely, orthogonal polarisation minimised the excitation (see Figure \ref{fig:PL}a for comparison of intensities). This is in agreement with the two-dimensional nature of the bright excitons, that offer maximum cross section for polarisation of light when it is parallel to the flake \cite{schuller2013}. By rotating the polarisation of the incident light, the excitonic emission is not completely extinguished. Assuming the finite theoretical limit of the extinction, we speculate it is further reduced due to the irregularities in the monolayer surface (introduced during the dry-stamping transfer process), which act as scattering centres, visible in Figure \ref{fig:panel}d. Similar photoluminescence curves are recorded for intermediate values of polarisation in the increments of \SI{10}{\degree}. The  resultant spectra are integrated, normalised for power (to mitigate the laser instabilities) and plotted against the polarisation angle (Figure \ref{fig:PL}b). Power for the normalisation is recorded after passing through the chip, at the position of a CCD sensor (Figure \ref{fig:setupPL}). The normalised points are then fitted to a sine function that demonstrates the polarisation dependence of photoluminescence with high extinction ratio of 31 between parallel and perpendicular states, which compares favourably with previous reports \cite{tan2014}.

In addition, we have performed measurements of photoluminescence at the end face of the chip comparing two identical waveguides, one of which contained the flake. The flake generates bright exciton emission at a normal to its plane, which does not couple back into the guided mode and we have indeed observed no resolvable photoluminescence signal at the output of the waveguide. Further to this, we aim to extend the experiment to investigate emission of the dark exciton in MoS$_2$ and other dichalcogenide monolayers \cite{robert2017,wang2017}, which has the potential for waveguide out-coupling in our geometry.

\subsection{Conclusion}

We have presented a working prototype of a photonic chip with an atomically thin MoS$_2$ flake, which is encapsulated directly within the single mode of guided laser light. The observed photoluminescence spectra are consistent with the ones present in the literature for monolayers of MoS$_2$. Sensitivity of the photoluminescence signal to polarisation of pump light is shown, with high extinction ratio and exact expected periodicity of \SI{180}{\degree}. The chip has shown robustness and ease of handling, being reused in multiple experiments over a period of months without any noticeable degradation in the MoS$_2$ flake. Such capabilities can enable further experiments dealing with optically active materials, such as 2D monolayers, nano-diamonds, crystalline and organic single photon emitters, and similar.

\subsection{Methods}

The elastomeric waveguide chip is fabricated out of polydimethylsiloxane (PDMS) in a two-step soft lithography process \cite{grieve2017}, with an additional step for material encapsulation. A lithographic mould is prepared by defining \SI{5}{\micro\metre} by \SI{1.6}{\micro\metre} waveguides on a silicon wafer using laser direct-write in the photoresist (AZ1512HS, AZ Electronic Materials). The PDMS liquid precursor is spun onto the mould and cured at \SI{150}{\celsius} to create a \SI{5}{\micro\metre} thick high-density layer. The MoS$_2$ flake is transferred to the cured PDMS layer at the location of a waveguide, using dry-stamping technique \cite{castellanos2014}. \SI{1}{\milli\metre} of PDMS liquid precursor is then poured on the top of the structure and cured at \SI{70}{\celsius} to form a low-density layer and provide a structural base for the chip. The chip is then peeled from the silicon substrate and end-faces are cut using a microtome blade to enable coupling of an optical fibre. Optical microscope images of the encapsulated flake are shown in Figure \ref{fig:panel}c to e.

MoS$_2$ was exfoliated from the bulk single-crystal flake (SPI) by the scotch-tape method onto the surface of the clean home-made PDMS slab. The monolayer was identified by contrast, photoluminescence and Raman spectroscopy. Selected flake was then encapsulated in the PDMS waveguide, as explained above.

\subsection{Funding statement}
G.E. acknowledges the Singapore National Research Foundation for funding the research under medium-sized centre programme. G.E. also acknowledges support from the Ministry of Education (MOE), Singapore, under AcRF Tier 2 (MOE2015-T2-2-123, MOE2017-T2-1-134) and AcRF Tier 1 (R-144-000-387-114). A.L., F.A., J.A.G. and K.F.N. acknowledge support from the Ministry of Education (MOE), Singapore under AcRF Tier 3 (MOE2012-T3-1-009) and the National Research Foundation, Prime Minister’s Office, Singapore under its Research Centres of Excellence programme.


\bibliographystyle{acs}
\bibliography{references}

\end{document}